\documentclass[pss]{wiley2sp} 
\usepackage{amsmath}

\begin{document}

\title{Low frequency spin dynamics in the quantum magnet copper pyrazine dinitrate}

\titlerunning{Low frequency spin dynamics in copper pyrazine dinitrate}

\author{%
  H. K\"{u}hne\textsuperscript{\Ast,\textsf{\bfseries 1,2}},
  M. G\"{u}nther\textsuperscript{\textsf{\bfseries 1}},
  S.Grossjohann\textsuperscript{\textsf{\bfseries 3}},
  W. Brenig\textsuperscript{\textsf{\bfseries 3}},
  F.J. Litterst\textsuperscript{\textsf{\bfseries 2}},
  A.P. Reyes\textsuperscript{\textsf{\bfseries 4}},
  P.L. Kuhns\textsuperscript{\textsf{\bfseries 4}},
  M.M. Turnbull\textsuperscript{\textsf{\bfseries 5}},
  C.P. Landee\textsuperscript{\textsf{\bfseries 5}},
  and H.-H. Klauss\textsuperscript{\textsf{\bfseries 1,2}}}

\authorrunning{H. K\"{u}hne et al.}

\mail{e-mail
  \textsf{Kuehne@physik.tu-dresden.de}, Phone:
  +49-351-46332404, Fax: +49-351-46337734}

\institute{%
  \textsuperscript{1}\,Institut f\"{u}r Festk\"{o}rperphysik, TU Dresden,
01069 Dresden, Germany\\
  \textsuperscript{2}\,Institut f\"{u}r Physik der Kondensierten Materie, TU
Braunschweig, 38106 Braunschweig, Germany\\
\textsuperscript{3}\,Institut f\"{u}r Theoretische Physik, TU
Braunschweig, 38106 Braunschweig, Germany\\
\textsuperscript{4}\,National High Magnetic Field Laboratory,
Tallahassee, Florida 32310, USA\\
\textsuperscript{5}\,Carlson School of Chemistry and Department of
Physics, Clark University, Worcester, Massachusetts 01610, USA}



\abstract{%
%
%
%
\abstcol{%
The S=1/2 antiferromagnetic Heisenberg chain exhibits a magnetic
field driven quantum critical point. We study the low frequency spin
dynamics in copper pyrazine dinitrate (CuPzN), a realization of this
model system of quantum magnetism, by means of $^{13}$C-NMR
spectroscopy. Measurements of the nuclear spin-lattice relaxation
rate $T_1^{-1}$ in the vicinity of the saturation field are compared
with quantum Monte Carlo calculations of the
}{%
dynamic structure factor. Both show a strong divergence of low
energy excitations at temperatures in the quantum regime. The
analysis of the anisotropic $T_1^{-1}$-rates and frequency shifts
allows one to disentangle the contributions from transverse and
longitudinal spin fluctuations for a selective study and to
determine the transfer of delocalized spin moments from copper to
the neighboring nitrogen atoms.

}}

%
%

 \maketitle    

\section{Introduction}
Low dimensional quantum magnets, e.g. spin chains or ladders, are
model systems for the study of quantum criticality
\cite{Watson2001,Lorenz2008,Honda1998,Zapf2006,Zvyagin2007,Manaka1998,Garlea2007,Takigawa1996,Chaboussant1997}.
For the isotropic S=1/2 antiferromagnetic Heisenberg chain (AFHC),
 nuclear magnetic resonance (NMR) has been used to
probe the slow spin fluctuations in the full magnetic phase diagram
of the AFHC, i.e., from the low to the high field limit and at
temperatures from the quantum regime ($k_B T\ll J$) up to the
classical regime ($k_B T\gg J$)\cite{Kuehne2009}. The Hamiltonian of
the AFHC in an external field reads:
\begin{equation}\label{eqn1}
H = \sum_{i} (J \mathbf{S}_i \cdot \mathbf{S}_{i+1} - g \mu_B
\mathbf{B} \cdot \mathbf{S}_i),
\end{equation}
where ${\bf S}_i$ are spin operators and $J$ is the exchange energy.
At $B_c=2J/(g\mu_B)$ the system exhibits a quantum critical phase
transition, where a single Ising triplet as lowest elementary
excitation crosses the ground state and the system switches from
complete polarization into a Luttinger liquid of deconfined spinons.
The condensation of low energy excitations is often referred to as
critical slowing down and gives rise to a divergence of slow spin
fluctuations.\\
In this paper we present a NMR study of the low frequency spin
response for various magnetic fields close to $B_c$ = 14.9 T
\cite{hammar} in the metalorganic AFHC copper pyrazine dinitrate
(CuPzN). We discuss
experimental data for fields from 2.0 to 14.8 T and temperatures
from 1.6 to 30 K. The dynamics are probed by the nuclear
spin-lattice relaxation rate $T_1^{-1}$ and compared with quantum
Monte Carlo (QMC) calculations. We also study the anisotropy of
contributions from transverse and longitudinal spin fluctuations to
$T_1^{-1}$ and relate them to the anisotropic NMR frequency shift in
the low field regime. The results are used to experimentally select
the transverse spin fluctuations in the present work and to evaluate
the spin density transfer from the copper ions to the neighboring
nitrogen atoms in the pyrazine ring.

\begin{figure}
\begin{center}
\includegraphics[width=0.9\columnwidth]{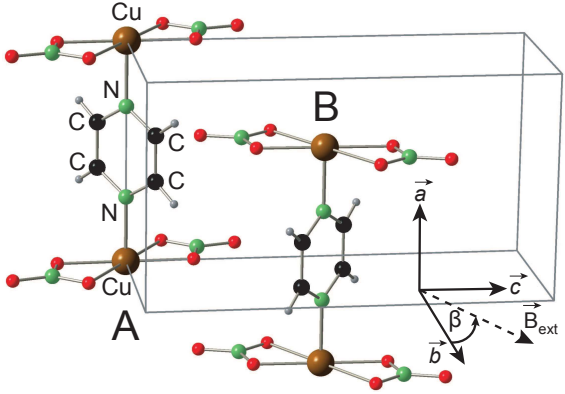}
\end{center}
\vskip -.5cm \caption[1]{Segments of two neighboring chains A and B
in the unit cell of CuPzN. The $^{13}$C atoms reside in the pyrazine
rings which are twisted against each other by $\simeq$ $80^\circ$ in
the $b$-$c$ plane.} \label{fig1} \vskip -.5cm
\end{figure}

\section{Experimental}
The compound CuPzN, i.e.,
Cu\-(C$_4$\-H$_4$\-N$_2$)\-(N\-O$_3$)$_2$ has been characterized by
inelastic neutron scattering, muon-spin relaxation, magnetothermal
transport, specific heat and magnetization measurements
\cite{hammar,lancaster,Stone03,solo}. All of these studies confirm
CuPzN to be one of the best realizations of the AFHC with a low
coupling constant $J/k_B=10.7$ K. This low value of $J$ allows the
study of the magnetic QCP in standard laboratory magnets. The
electronic $S = $1/2 spins are mainly located in the Cu$^{2+}$ 3d-
orbitals, whereas the $^{13}$C nuclei in the pyrazine ring were used
as $I$ = 1/2 NMR probes. The onsite copper nuclei exhibit an extremely
short spin-correlation time that is below the dead time of a
standard NMR spectrometer. The measurements were performed in an 8 
and a 17 T superconducting magnet with a modified Bruker CXP200
spectrometer and a home-built spectrometer, respectively. In both
cases a standard inversion-recovery spin-echo pulse sequence was
used.

\section{Nuclear Magnetic Resonance}
The nuclear spin-lattice relaxation rate $T_1^{-1}$ measures the
electronic spin fluctuations at the nuclear Larmor frequency
$\omega_n$. For the AFHC it can be written as~\cite{mor}
\begin{eqnarray}
\frac{1}{T_1}= \frac{\gamma_n^2}{2} \sum_{q} \sum_{\beta = x,y,z} [A
_{x \beta}^2 (q) + A _{y \beta}^2 (q)]\hspace{1.5cm} \nonumber
\\ \times \int_{- \infty} ^{\infty} < S_{\beta}
(q,t)S_{\beta} (-q,0)> e^{-i \omega_n t} dt
\label{eqn2} \\
=\frac{\gamma_n^2}{2} \sum_{q}[F_{\perp}(q) S_{\perp}(q, \omega_n) +
F_{z}(q) S_{z}(q, \omega_n)]. \label{eqn3}
\end{eqnarray}
Here, $F_{\perp}(q) = \sum_{\beta = x,y} [A _{x \beta}^2 (q) + A _{y
\beta}^2 (q)]$ and $F_{z}(q) =  A _{x z}^2 (q) + A _{y z}^2 (q)$ are
the geometrical form factors and $S_{\perp}(q, \omega)$ and
$S_{z}(q, \omega)$ are the dynamical structure factors of the
electronic spin system. $A_{\alpha \beta}$ with $\alpha,
\beta=x,y,z$ are the components of the hyperfine
coupling tensor $\underline{\mathbf{A}}(q)$.\\

\subsection{Anisotropy of the NMR shift $\delta$}

\begin{figure*}[htb]%
\subfloat[$B=2.0$ T, $T=7$ K.]{%
\includegraphics*[width=0.66\columnwidth]{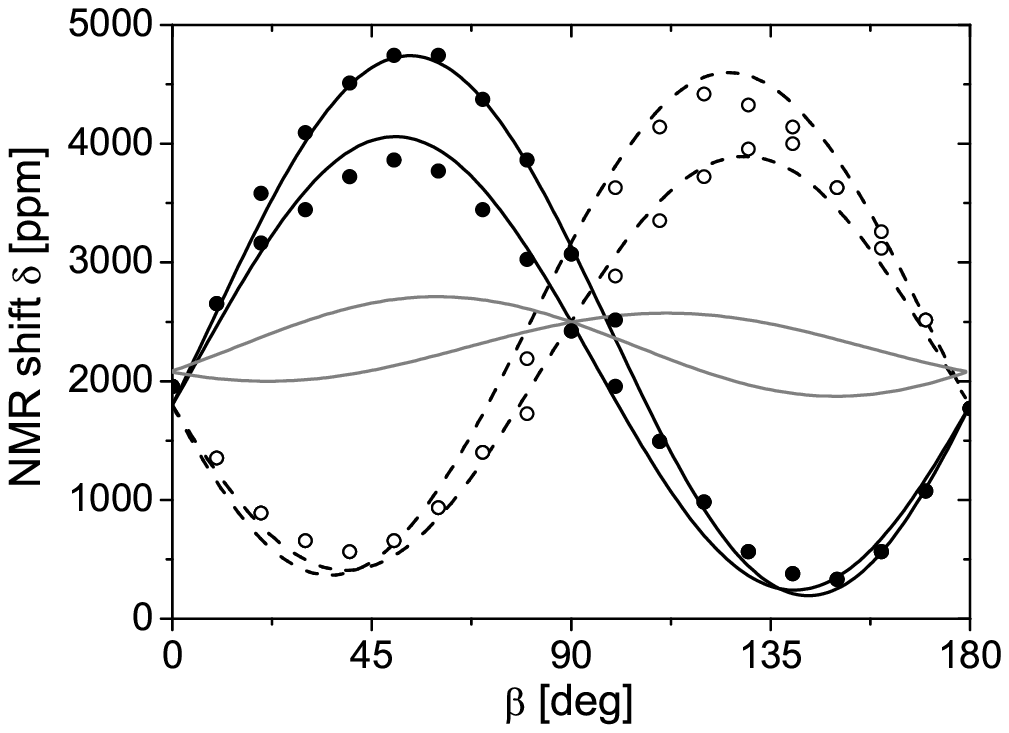}}\hfill
\subfloat[$B=2.0$ T, $T=7$ K.]{%
\includegraphics*[width=0.66\columnwidth]{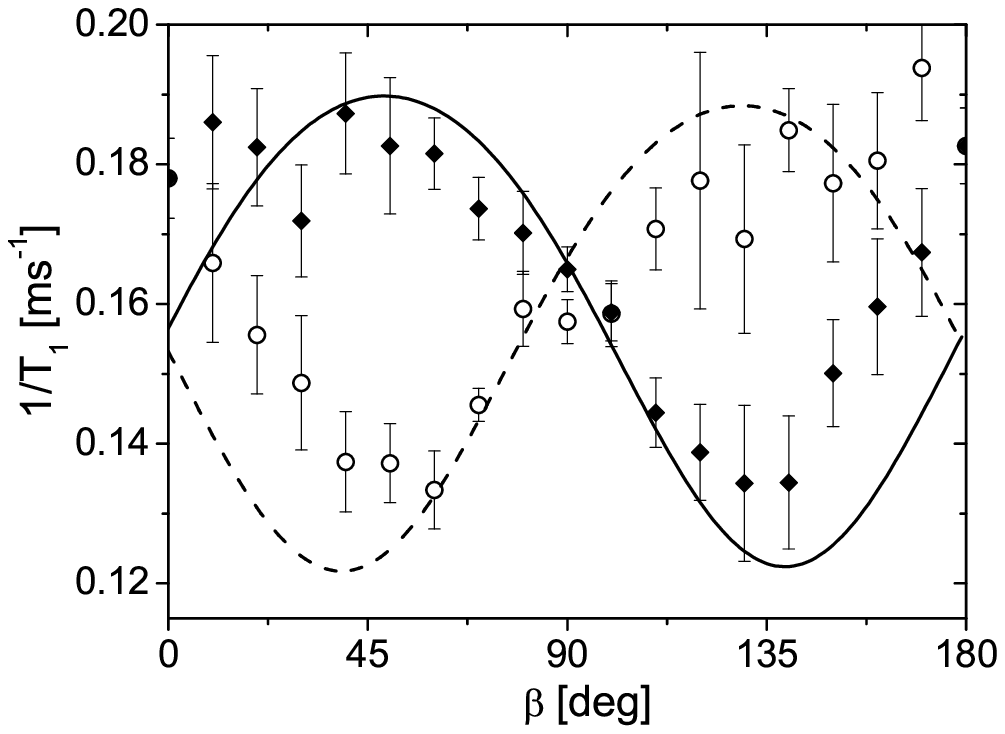}}\hfill
\subfloat[$B=2.0$ T, $T=7$ K.]{%
\includegraphics*[width=0.66\columnwidth]{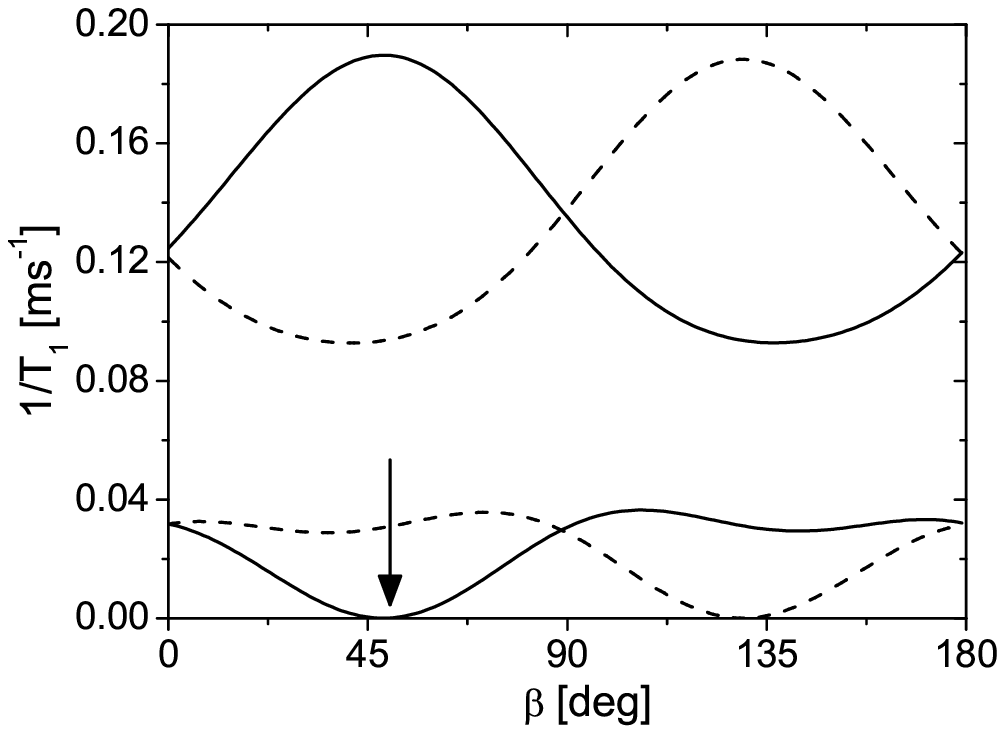}}\hfill
\caption{(a) NMR shift $\delta$ versus rotation angle $\beta$. As in
(b) and (c), full (line set A) and dashed (line set B) lines
correspond to calculations with a spin moment transfer of 10$\%$ in
the simulation, gray lines correspond to a transfer of 0$\%$. (b)
Experimental $T_1^{-1}$-rates versus $\beta$ compared with a
modulation of $T_1^{-1}$ based on Eqs. (\ref{eqn3}) and
(\ref{eqnshift}). (c) The modulation shown in (b) is disentangled
into rates that stem from transverse (large values) and longitudinal
(small values) fluctuations of the electronic moments. The arrow
marks $\beta = 50^\circ $, where the latter rate of set A goes to
zero.} \label{threesubfigures}
\end{figure*}

In CuPzN both an isotropic hyperfine part
$\underline{\mathbf{A}}_{iso}(q)$ and an anisotropic dipolar part
$\underline{\mathbf{A}}_{dip}(q)$ contribute to the coupling of the
$^{13}$C nuclei to the magnetic moments of the Cu$^{2+}$ electrons.
The NMR shift $\delta$ is proportional to the projection of the hyperfine
coupling tensor $\underline{\mathbf{A}}(q=0)$ on the direction of
the external field, i.e., the component $A_{zz} (0)$. Thus, for a
rotation of the crystal in a fixed external field the shift can be
extracted from:
\begin{eqnarray}
\hat{\underline{\mathbf{{A}}}}(0)= \underline{\mathbf{D}} \cdot ((
\underline{\mathbf{A}}_{dip}(0) + \underline{\mathbf{A}}_{iso}(0))
\cdot \underline{\mathbf{\chi_{el}}} +
\underline{\mathbf{\sigma}})\cdot \underline{\mathbf{D}}^{-1},
\label{eqnshift}
\end{eqnarray}
where $\underline{\mathbf{\chi_{el}}}$ is the tensor of the
electronic susceptibility, $\underline{\mathbf{\sigma}}$ is the
diamagnetic tensor, and $\underline{\mathbf{D}}$ is a rotation
matrix. Fig. \ref{threesubfigures}(a) shows the experimental shift
$\delta$ ($\beta$) versus a simulation based on localized dipole
moments for a rotation around the crystallographic $a$-axis, where
$\beta=0^\circ$ corresponds to an external field parallel to the
$b$-axis, compare Fig. \ref{fig1}. The assumption that 100$\%$ of
the spin magnetic moment resides on the Cu sites cannot describe
the experimental data. A very good agreement is achieved for a spin
moment transfer of $10 \%$ from copper to each of the neighboring
nitrogen atoms in the pyrazine ring and a misalignment of $5^\circ$
between the actual rotation axis and the crystallographic $a$-axis.
The transfer of delocalized spin moments was observed in similar
systems \cite{wolter}. For the resulting isotropic hyperfine part we
find $\delta_{iso}=3600$ ppm. The existence of two sets of spectral
lines results from the relative twisting of pyrazine rings on two
neighboring chains by $\simeq$ $80^\circ$. According to Fig.
\ref{fig1},
those will be labeled as line set A and B in the following.\\

\subsection{Transverse dynamic structure factor}
Before turning to the $T_1^{-1}$ data, we present our method of
calculation for the field and temperature-dependent transverse
dynamic structure factor $S_{\perp}(q,\omega_n)$. Switching to
imaginary time $\tau$ the latter reads  $S_{\perp}(q,
\tau)=\frac{1}{\pi}\int_{0}^{\infty}d\omega S_{\perp}(q,
\omega)K(\omega, \tau)$, with a kernel $K(\omega,
\tau)=e^{-\tau\omega} + e^{-(\beta-\tau)\omega}$ and $\beta=1/T$. In
real-space $S_{\perp}(q,\tau)$ can be calculated efficiently, using
QMC. Following Ref. \cite{Sandvik1992}
\begin{eqnarray}\label{eqn5}
S_{i, j}\left( \tau\right) =\left\langle \sum_{p,m=0}^{n}
\frac{\tau^m(\beta-\tau)^{n-m}n!}{\beta^n(n+1)(n-m)!m!} \times
\right . \phantom{a a a}
&&\nonumber\\
\left . S^{+}_{i}(p)S^{-}_{j}(p+m) \right\rangle_W ~, &&
\end{eqnarray}
where $S_{\perp}(q, \tau)=\sum_a e^{iq a}S_{a,0}(\tau)/N$ and $a,0$
label lattice sites in a chain of length $N$.
$\langle\ldots\rangle_W$ refers to the Metropolis weight of an
operator string of length $n$ generated by the stochastic series
expansion of the partition function
\cite{Sandvik1999a,Syljuaasen2002}, and $p,m$ are positions in this
string.

Analytic continuation from imaginary times, i.e., $S_{\perp}(q,
\tau)$, to real frequencies, i.e., $S_{\perp}(q, \omega)$, is
performed by the maximum entropy method (MaxEnt), minimizing the
functional $Q=\chi^2/2 - \alpha\sigma$ \cite{bryan1,Jarrell1996}.
Here $\chi$ refers to the covariance of the QMC data to the MaxEnt
trial-spectrum $S_{\alpha\perp}(q, \omega)$. Overfitting is
prevented by the entropy $\sigma = \sum_{\omega}S_{\alpha\perp}(q,
\omega)\ln[S_{\alpha\perp}(q, \omega)/m(\omega)]$. We have used a
flat default model $m(\omega)$, matching the zeroth moment of the
trial spectrum. The optimal spectrum follows from the weighted
average
\begin{equation}
S_{\perp}(q, \omega) = \int_{\alpha}d\alpha P[\alpha | S(q,
\tau)]S_{\alpha\perp}(q, \omega) ~,
\end{equation}
with the probability distribution $P[\alpha | S(q, \tau)]$ adopted
from Ref. \cite{bryan1}.\\

\subsection{Anisotropy of spin fluctuations}
\begin{figure}[b]
\begin{center}
\includegraphics[width=1.0\columnwidth]{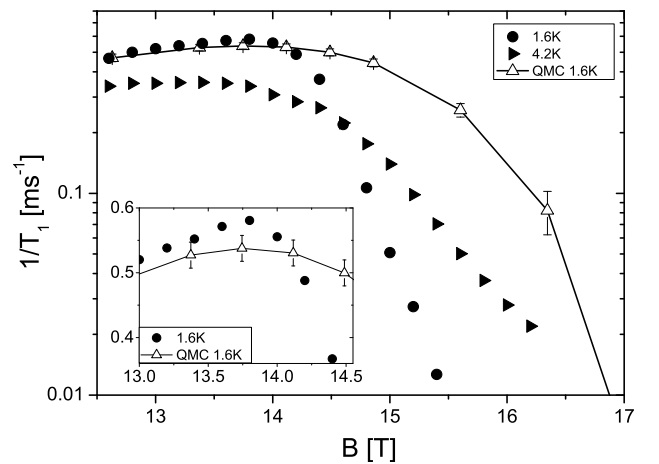}
\end{center}
\vskip -.5cm \caption[1]{Field dependence of the nuclear
spin-lattice relaxation rate of $^{13}$C close to the critical
field. The exponential decrease of data reflects the linear opening
of the spin gap with field. The inset highlights the maximum of
$T_1^{-1}(B)$ at $B=\tilde{B}_c$ and $T=1.6$ K. All solid lines are
a guide to the eye.} \label{fig3} \vskip -.5cm
\end{figure}

The ratio of longitudinal and transverse spin fluctuations
contributing to the total spin-lattice relaxation rate at a given
field can be evaluated from the anisotropy of $T_1^{-1}$
\cite{azevedo}. Our previous studies show that in this system the
geometrical form factors for the case of $^{13}$C-NMR are almost
q-independent ~\cite{Kuehne2009}. Therefore, the spin-lattice
relaxation rate only depends on $S_{\perp}(q)$, $S_{z}(q)$ and the
orientation between the external field and the local hyperfine
fields, which in turn are well known from the study of $\delta$
($\beta$). Fig. \ref{threesubfigures} (b) shows the experimental
$T_1^{-1}$-rates versus a modulation of the relaxation rates based
on equations (\ref{eqn3}), (\ref{eqnshift}) and an almost isotropic
dynamic structure factor calculated by QMC. The modulation was
fitted to the experimental data by a single scaling factor
determined by a least-squares fit. We find a good agreement for the
comparison between our experimental data and the quantum Monte Carlo
results. In this work we want to compare the experimentally and
theoretically determined field- and temperature-dependent
$T_1^{-1}$-rates stemming from transverse fluctuations only.
Therefore, the contribution from longitudinal fluctuations $S_{z}(q,
\omega)$ was minimized by minimizing $F_{z}(q)$. Fig.
\ref{threesubfigures} (c) shows that for line set A the longitudinal
fluctuation contribution to $T_{1}^{-1}$ approaches zero for $\beta
=50^\circ$. This line set and orientation was used for the
subsequent measurements shown in Figs. \ref{fig3} and \ref{fig4}.
Note that the transverse spin fluctuations also dominate the nuclear
relaxation due to the large Fermi contact term in $F_{\perp}(q)$.\\

\subsection{$\mathbf{1/T_1}$: field dependence}
\begin{figure}
\begin{center}
\includegraphics[width=1.0\columnwidth]{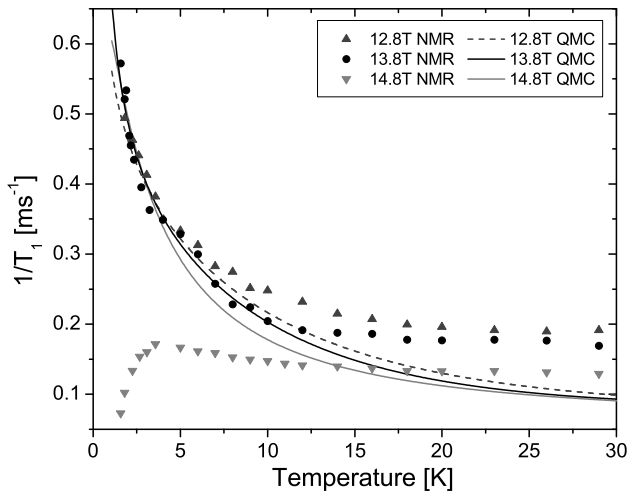}
\end{center}
\vskip -.5cm \caption[1]{Temperature dependence of the nuclear
spin-lattice relaxation rate of $^{13}$C for different external
fields in the vicinity of the quantum critical point. All QMC data
lines are polynomial fits to analytic continuations of a 128 site
system. The theoretical errors range from $10\%$ at low temperatures
to $20\%$ at high temperatures, estimated from different MaxEnt
regularization parameters.} \label{fig4} \vskip -.5cm
\end{figure}

In Fig. \ref{fig3} we compare the observed NMR rates with the QMC
results versus magnetic field in the quantum regime $k_BT\ll J$,
with $T=1.6$ K. The QMC data shown in Figs. \ref{fig3} and
\ref{fig4} were calculated from the transverse dynamic structure
factor $S_{\perp}(q,\omega)$ \cite{Grossjohann2009} and scaled with
a factor assigned at 2.0 T and high temperatures. The similarity
between experiment and theory is remarkable. For both we find a
pronounced maximum of $T_1^{-1}(B)$ at $B=13.8$ T shifting to lower
fields with increasing temperature. To interpret these results, we
note that in the fully polarized state for $B>B_c$, single magnons
are exact eigenstates of Eq. (\ref{eqn1}) with a dispersion of
\begin{eqnarray}
E_>(k) = J \cos (k) + g \mu_B B~.
\end{eqnarray}
$E_>(k)$ displays a field-driven excitation gap of $g \mu_B B-J$
leading to an exponential decrease in $T_1^{-1}(B)$ at fixed $T$ and
for $B>B_c$. The rates calculated by QMC display a broader maximum
than the measured data, but drop with the same slope for fields
above $16$ T. At $B=B_c$ the dispersion touches the zero at
$k=\pi/2$ with a {\it quadratic} momentum dependence yielding a
van-Hove type of critical DOS. This leads to the maximum in
$T_1^{-1}$, tending to diverge as $T\rightarrow 0$. For both, NMR
experiment and QMC, the maximum in Fig. \ref{fig3} occurs at
$\tilde{B}_c\approx 13.8 $ T, which is slightly less than the
saturation field of $B_c=14.9$ T for the magnetization. This
downshift is most likely a finite- temperature effect of excitations
populating the gap.\\

\subsection{$\mathbf{1/T_1}$: temperature dependence}
In Fig. \ref{fig4} the temperature dependent $T_1^{-1}$ rates for
fields in the vicinity of the saturation field are shown. We find
the strongest low temperature divergence at $\tilde{B}_c$=13.8 T.
This is very suggestive of critical scattering as $T\rightarrow 0$.
With increasing temperature the van-Hove singularity in the DOS at
$B_c$ is smeared leading to the decrease in $T_1^{-1}$. At $B_c
\simeq B = 14.8$ T the experimental data show no divergence but
rather the opening of a spin gap. Possibly this field is slightly
above the true saturation field for this orientation due to the
sensitivity of the exact value of $B_c$ to the anisotropic g-factor.
The QMC data for 12.8 and 13.8 T show a satisfying agreement with
the experiment, with the same divergence at low temperatures. In
contrast to the experimental data, at 14.8 T there is still a
divergence as the quantum regime
is approached.\\
In the classical regime $k_BT\gg J$ the rate is decreasing with
increasing fields. This is indicative of an excitation spectrum
dominated by spin-diffusion modes from $q=0$.
%

\section{Conclusion}

We performed a complementary experimental and theoretical study of
the low-frequency spin dynamics in the isotropic S=1/2
antiferromagnetic Heisenberg chain system copper pyrazine dinitrate.
We find clear evidence for critical dynamics close to a
field-induced QCP, as probed by the nuclear spin-lattice relaxation
rate $T_1^{-1}$. Experimental data and QMC calculations are in good
agreement and show a pronounced maximum in $T_1^{-1}$ in the
vicinity of the saturation field. In this study, the anisotropic
contribution from longitudinal spin fluctuations to $T_1^{-1}$ was
minimized. From the analysis of the NMR frequency shift we conclude
a spin moment transfer of $10\%$ from copper to each of the
neighboring nitrogen atoms in the pyrazine ring.

\begin{acknowledgement}
Part of this work was performed at the National High Magnetic Field
Laboratory, supported by NSF Cooperative under Agreement No.
DMR-0084173, by the State of Florida, by the DOE and the DFG under
Grants No. KL1086/6-2 and No. KL1086/8-1. One of us (W.B.)
acknowledges partial support by the DFG through Grants No. BR
1084/4-1 and BR 1084/6-1. He also thanks the KITP for hospitality,
where this research was supported in part by the NSF under Grant No.
PHY05-51164. \vskip -.5cm
\end{acknowledgement}

%
%

\end{document}